\documentclass[twocolumn]{aastex631}

\usepackage{natbib}
\usepackage{graphicx}
\usepackage{graphics}
\usepackage{amsmath}
\usepackage{color}
\usepackage{hyperref}
\definecolor{lblue}{rgb}{0,0.2,0.6}
\definecolor{dgreen}{rgb}{0.1,0.6,0.3}
\usepackage{enumitem}
\usepackage{nameref}
\usepackage{varioref}
\usepackage{hyperref}
\usepackage{cleveref}
\usepackage{amsmath,amssymb,amsxtra,amsfonts}   % to use pmatrix, etc.
\usepackage{txfonts}

\newcommand       \be           {\begin{equation}}
\newcommand       \ee           {\end{equation}}

\begin{document}

\title{The UV luminosity function at $1.6<\lowercase{z}<2.2$ from UVCANDELS}

\correspondingauthor{Lei Sun, Xin Wang}
\email{sunl@ucas.ac.cn \quad xwang@ucas.ac.cn}

%% lead authors
\author[0009-0004-6325-7839]{Lei Sun}
\affil{School of Astronomy and Space Science, University of Chinese Academy of Sciences, Beijing 100049, China}

\author[0009-0003-1463-4397]{Shamuhawu Hailanhazi}
\affil{School of Astronomy and Space Science, University of Chinese Academy of Sciences, Beijing 100049, China}

\author[0000-0002-9373-3865]{Xin Wang}
\affil{School of Astronomy and Space Science, University of Chinese Academy of Sciences, Beijing 100049, China}
\affil{Institute for Frontiers in Astronomy and Astrophysics, Beijing Normal University,  Beijing 102206, China}
\affil{National Astronomical Observatories, Chinese Academy of Sciences, Beijing 100101, China}

\author[0000-0002-7064-5424]{Harry I. Teplitz}
\affil{IPAC, Mail Code 314-6, California Institute of Technology, 1200 E. California Blvd., Pasadena CA, 91125, USA}

\author[0000-0001-7166-6035]{Vihang Mehta}
\affiliation{IPAC, Mail Code 314-6, California Institute of Technology, 1200 E. California Blvd., Pasadena CA, 91125, USA}

\author[0000-0002-8630-6435]{Anahita Alavi}
\affiliation{IPAC, Mail Code 314-6, California Institute of Technology, 1200 E. California Blvd., Pasadena CA, 91125, USA}

%% coauthors providing comments
\author[0000-0003-1949-7638]{Christopher J. Conselice}
\affiliation{Jodrell Bank Centre for Astrophysics, University of Manchester, Oxford Road, Manchester, UK}

\author[0000-0001-8587-218X]{Matthew Hayes}
\affiliation{Stockholm University, Department of Astronomy and Oskar Klein Centre for Cosmoparticle Physics, AlbaNova University Centre, SE-10691, Stockholm, Sweden}

\author[0000-0003-1581-7825]{Ray A. Lucas}
\affiliation{Space Telescope Science Institute, Baltimore, MD 21218, USA}

\author[0000-0002-9946-4731]{Marc Rafelski}
\affiliation{Space Telescope Science Institute, Baltimore, MD 21218, USA}
\affiliation{Department of Physics and Astronomy, Johns Hopkins University, Baltimore, MD 21218, USA}

\author[0000-0002-4935-9511]{Brian Siana}
\affiliation{Department of Physics and Astronomy, University of California, Riverside, Riverside, CA 92521, USA}

\author[0000-0001-8156-6281]{Rogier A. Windhorst}
\affiliation{School of Earth and Space Exploration, Arizona State University,Tempe, AZ 85287-6004, USA}

\author[0000-0003-3466-035X]{L. Y. Aaron Yung}
\affiliation{Space Telescope Science Institute, Baltimore, MD 21218, USA}

%% other UVCANDELS builders
\author[0000-0001-9440-8872]{Norman Grogin}
\affiliation{Space Telescope Science Institute, Baltimore, MD 21218, USA}

\author[0000-0003-3759-8707]{Ben Sunnquist}
\affiliation{Space Telescope Science Institute, Baltimore, MD 21218, USA}

%\author{the UVCANDELS team}

\begin{abstract}

We present an analysis of the rest-frame UV luminosity function (LF) of galaxies at $1.6<z<2.2$ using the UVCANDELS F435W imaging data. Restricting our analysis to sources above $30\%$ completeness to minimize uncertainties at the faint end, we construct a catalog of 5,212 galaxies at $1.6<z<2.2$ down to an absolute magnitude of $M_{\rm UV} = -16.8$, using well-tested photometric redshifts. We perform a maximum likelihood estimate to derive the best-fit Schechter parameters of the UV LF at a median redshift of $z\sim1.9$. We report a best-fit faint-end slope of $\alpha = -1.586 \pm 0.042$. Splitting the sample into two redshift bins, we obtain $\alpha=-1.488 \pm 0.055$ at $z\sim1.75$ and $\alpha=-1.604 \pm 0.066$ at $z\sim2.05$ ($\sim1.4\sigma$ difference), alongside a $\sim2.3\sigma$ brightening in $M^{*}$. Integrating our best-fit LF down to $M_{\rm UV} = -15$, we derive $\rho_{\rm UV} = (1.821 \pm 0.037) \times 10^{26}\ \rm erg\ s^{-1}\ Hz^{-1}\ Mpc^{-3}$ at $z\sim1.9$. Combining our F435W measurements with previous UVCANDELS F275W results and other literature data, we model the redshift evolution of $\rho_{\rm UV}$ from $z\sim0$ to $z\sim10$.

\end{abstract}

\keywords{Galaxies (573); Galaxy evolution (594); Luminosity function (942); High-redshift galaxies (734)}

\section{Introduction}
\label{sec:intro}

The rest-frame UV luminosity function (LF) is a fundamental tool for tracing galaxy formation and evolution across cosmic time. In recent years, the UV LF has been measured extensively at both low redshifts ($z\lesssim1$) using GALEX, AstroSat, XMM-OM, Swift UVOT and HST \citep[e.g.,][]{Arnouts05, Sharma.2024, Bhattacharya.2025, Sun.2026}, and at high redshifts ($z\gtrsim6$) where JWST observations are now rapidly delivering abundant measurements, probing galaxies from the reionization epoch to the cosmic dawn \citep[e.g.,][]{Donnan.2024, Adak.2024, Willott.2024, Robertson.2024, Bagley.2024, Mcleod.2024, Perez.2025, Whitler.2025}. 

At the intermediate epoch of ``cosmic noon'' ($1.5\lesssim z\lesssim3$), the peak of cosmic star formation history, there are also numerous UV LF measurements, including the canonical LBG survey from \cite{Reddy.2009} covering 31 independent fields, VVDS \citep{Cucciati12}, UVUDF \citep{Mehta.2017}, HFF lensing fields \citep{Alavi16}, CLAUDS/HSC-SSP \citep{Moutard20}, and the comprehensive compilation from HST legacy fields \citep{Bouwens.2021, Bouwens.2022c}. Despite this wealth of data, a systematic, deep, and large-area survey with uniform selection and processing is needed to further refine the UV LF parameters at these redshifts.

The Ultraviolet Imaging of the Cosmic Assembly Near-infrared Deep Extragalactic Legacy Survey Fields (UVCANDELS; HST-GO-15647; PI: Teplitz; \citealt{Wang.2025}) provides an ideal opportunity to further advance the UV LF measurements at the intermediate redshifts. As a Cycle-26 HST Treasury Survey, UVCANDELS obtained deep UV imaging with WFC3/F275W over $\sim426$ arcmin$^2$ across four CANDELS fields and parallel optical imaging with ACS/F435W over $\sim289$ arcmin$^2$ in the COSMOS and EGS fields, reaching $5\sigma$ depths of 27 ABmag (F275W) and 28 ABmag (F435W) in $0''.2$ apertures. In the first paper of the UVCANDELS UV LF series (\citealt{Sun.2024}; hereafter Paper I), we used the primary F275W channel to measure the UV LF at $0.6<z<1$ ($z\sim0.8$), reaching $M_{\mathrm{UV}}=-15.5$ and obtaining a faint-end slope of $\alpha=-1.359\pm0.041$. The parallel F435W data are ideally suited to extend this work to cosmic noon, where the observed F435W band directly samples the rest-frame 1500~\AA\ emission at $1.6<z<2.2$. Compared to previous surveys at these redshifts, the UVCANDELS F435W data bridge the gap between the small-area, ultra-deep HST fields and the wide but shallower ground-based surveys, with a combination of $\sim289$ arcmin$^2$ area and 28 ABmag depth that enables a robust and homogeneous measurement of the UV LF parameters. In this second paper, we apply the same data reduction, photometry, completeness simulations, and LF fitting framework as Paper I to the F435W imaging, constructing a catalog of 5,212 galaxies down to $M_{\mathrm{UV}}=-16.8$ and measuring $\alpha$ and $\rho_{\mathrm{UV}}$ at $z\sim2$. By combining these new measurements with Paper I and literature data, we trace the continuous evolution of $\alpha$ and $\rho_{\mathrm{UV}}$ from $z\sim0$ through cosmic noon and beyond.

The structure of this paper is as follows: in Section~\ref{sec:data_phot}, we briefly summarize the UVCANDELS observations, data reduction, and photometric catalogs; Section~\ref{sec:sample} presents our sample selection and contamination removal; Section~\ref{sec:comp} describes our completeness simulations and corrections; Section~\ref{sec:lf} details the maximum likelihood estimation of the UV LF and presents the best-fit Schechter parameters; Section~\ref{sec:discuss} discusses the redshift evolution of the LF parameters, the UV luminosity density and the modeling of its redshift evolution; and we summarize our conclusions in Section~\ref{sec:sum}.

Throughout this paper, we adopt a flat $\Lambda$CDM cosmology with $\Omega_m=0.3$, $\Omega_\lambda = 0.7$ and $H_0=70$ km s$^{-1}$ Mpc$^{-1}$. All magnitudes used are AB magnitudes \citep{Oke83}.

\section{Data and photometry}
\label{sec:data_phot}

The UVCANDELS F435W imaging data, obtained in parallel with the primary F275W observations, were reduced following the procedures detailed in \cite{Wang.2025}. The observations reach a $5\sigma$ depth of $28$~ABmag in a $0''.2$ aperture, covering a total area of $\sim289$~arcmin$^2$ in the COSMOS and EGS fields (the two fields where UVCANDELS provides new F435W data). The F435W mosaics are astrometrically aligned to the world coordinate system used by the CANDELS images \citep{Koekemoer.2011, Grogin.2011}.

We adopt the UV-optimized aperture photometry method developed in Paper I. Briefly, we use F606W as the detection image to define object isophotes and perform photometry on the original F435W mosaics without point-spread function (PSF) homogenization (the ``V-isophot'' method). This approach maximizes the signal-to-noise ratio (S/N) for F435W-detected sources. For the UVCANDELS F275W data, Paper I demonstrated a factor of $\sim1.5\times$ improvement over the traditional H-isophot method; given the comparable PSF mismatch between F435W and F160W, this approach is also expected to benefit the F435W photometry, particularly for faint galaxies. The same SExtractor \citep{Bertin.1996} parameters as in Paper I (Table~1 therein) are applied. The F435W photometric catalog has been made publicly available on the Barbara A. Mikulski Archive for Space Telescopes (MAST) by \citet{UVCANDELS_MAST}\footnote{The UVCANDELS High Level Science
Product is available at \url{https://archive.stsci.edu/hlsp/uvcandels} with\dataset[DOI: 10.17909/8s31-f778]{https://doi.org/10.17909/8s31-f778}}.

\section{Sample selection}
\label{sec:sample}

We select galaxies based on photometric redshifts (photo-z) derived from the full UV-to-IR multiwavelength catalogs, which include the new UVCANDELS F435W data presented in this work. The photo-z measurements are taken from \cite{Mehta.2024}, who performed a detailed analysis (including PSF homogenization to the F160W band) using a combination of codes including EAZY \citep{Brammer.2008}, BPZ \citep{Benitez00, Coe06}, LePhare \citep{Arnouts99, Ilbert06}, and zphot \citep{Giallongo98, Fontana00}, providing full redshift probability distributions for over 150,000 sources across the four CANDELS fields. The redshift range for our analysis is chosen as $1.6 < z < 2.2$. This selection ensures that the rest-frame $1500$~\AA\ emission falls within the high-transmission region of the F435W filter (i.e., within its FWHM bandpass), thereby minimizing the K-correction uncertainty. We require $\mathrm{S/N} \geq 3$ in the F435W band to ensure reliable flux measurements. Within this target redshift range, we find a photometric redshift accuracy of $\sigma_{\mathrm{NMAD}} = 0.0417$ with an outlier fraction of $\eta = 1.72\%$.

To ensure the purity of our sample, we carefully remove potential contaminants. We cross-match our sample with the CANDELS AGN catalogs based on X-ray observations, compiled by D. Kocevski et al. (private communication), and remove all matched sources. The number of X-ray sources rejected varies across the two fields, due to the different depths of the X-ray observations and survey areas. We also implement star/galaxy separation using the UVCANDELS F606W catalog (the detection band in our V-isophot photometry). Sources are identified as stars and rejected if they simultaneously satisfy all of the following criteria: $\mathrm{S/N} > 3$, $\mathrm{CLASS\_STAR} > 0.9$, elongation $= \mathrm{A\_IMAGE} / \mathrm{B\_IMAGE} < 1.2$, $f_{3\mathrm{arcsec}} / f_{0.7\mathrm{arcsec}} < 1.3$, and $m_{\mathrm{F606W}} < 22$~ABmag. Finally, at the bright end ($M_{\mathrm{UV}} < -20$), we perform a visual inspection to exclude any remaining spurious objects, such as those with incorrect segmentation.

After removing contaminants, we select a total of 6,254 galaxies that satisfy the redshift and S/N criteria. Their rest-frame absolute magnitude at $1500$~\AA\ is then computed as:

\begin{equation}
M_{1500} = m_{\mathrm{F435W}} - 5\log\left(\frac{d_L}{10~\mathrm{pc}}\right) - K_{\mathrm{cor}},
\label{eq:m1500}
\end{equation}

\noindent where $d_L$ is the luminosity distance at the galaxy redshift, and $K_{\mathrm{cor}}$ represents the K-correction from the emission wave band centered around rest-frame $1500$~\AA\ to the observed F435W band. For each galaxy, we estimate $K_{\mathrm{cor}}$ via interpolation on a precomputed $K_{\mathrm{cor}}(z,\beta)$ table, assuming a power-law spectrum $F(\lambda)\ \propto\ \lambda^{\beta}$ with a cutoff at $1216$~\AA\ (i.e., flux set to zero blueward of Ly$\alpha$) to approximate the effect of IGM absorption \citep{Hogg02}. The UV continuum slope $\beta$ for each galaxy is obtained from the SED fitting results \citep{Mehta.2024}. The F435W FWHM ($3879$--$4818$~\AA) corresponds to rest-frame ranges of $\sim1490$--$1850$~\AA\ at $z=1.6$ and $\sim1210$--$1510$~\AA\ at $z=2.2$ (bandwidths of $\sim360$~\AA\ and $\sim290$~\AA, respectively); across our redshift range ($1.6<z<2.2$), the bandpass therefore samples predominantly redward of Ly$\alpha$, dominated by the intrinsic UV continuum at $\sim1500$~\AA. We therefore simply adopt the $1216$~\AA\ cutoff model for the $K$-correction. The resulting $K$-correction is typically $\lesssim0.1$--$0.2$ mag over the range of $\beta$ values in our sample, and is therefore subdominant compared to the photometric and redshift measurement uncertainties. The derived $M_{1500}$ values are presented without dust attenuation correction.

\begin{figure}
\centering
\includegraphics[width=0.45\textwidth]{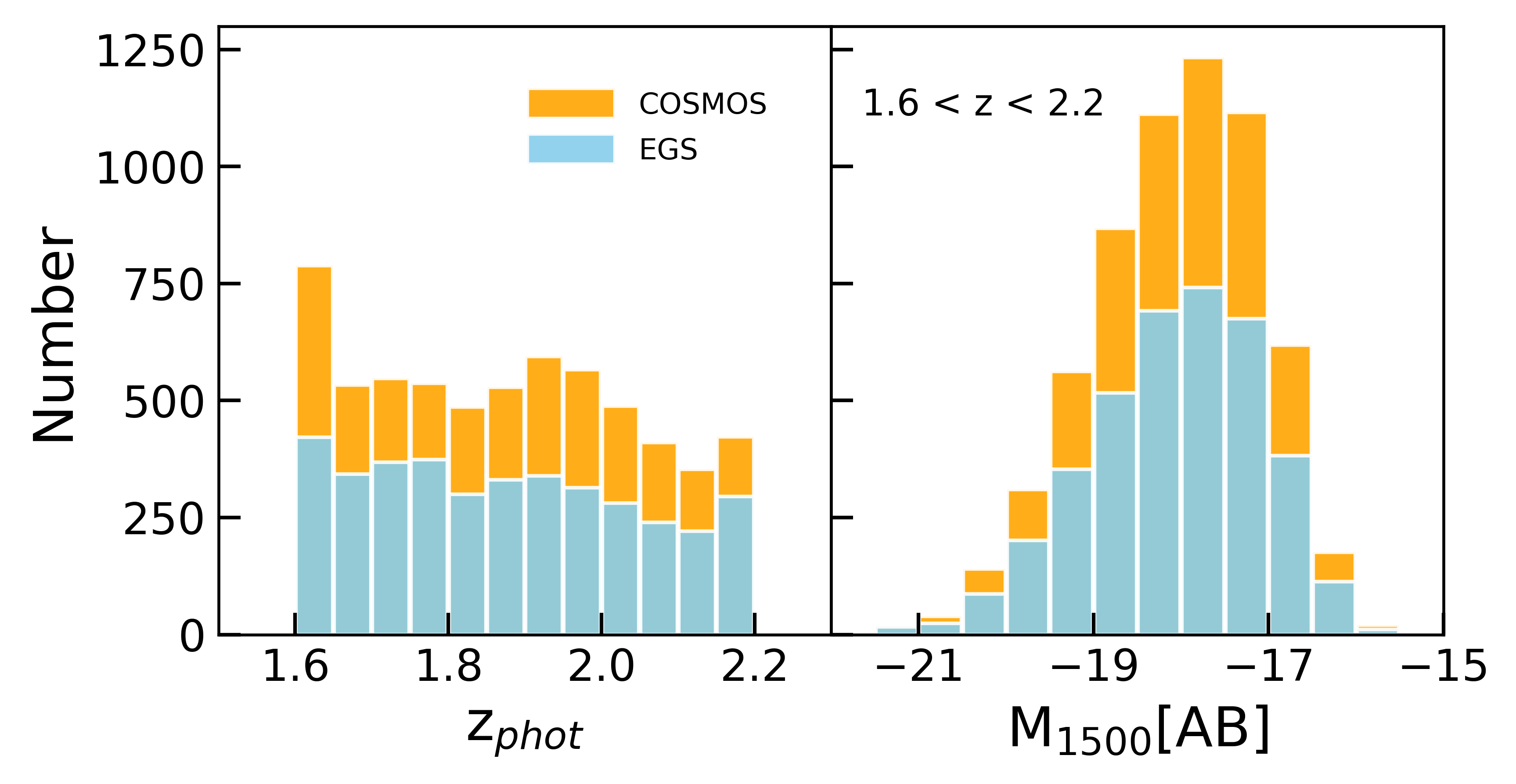}
\caption{The photometric redshift (left panel) and rest-frame 1500 \AA\ absolute magnitude (right panel, measured from F435W flux without dust attenuation correction) distributions of the selected sources from the UVCANDELS photometry catalogs at $1.6<z_{\rm phot}<2.2$ that have ${\rm SNR}\geq3$ for F435W flux. The faintest magnitude reaches $-15.5$. Possible contamination from stars, AGNs and spurious objects have been examined and excluded.}
\label{fig:M1500}
\end{figure}

Figure~\ref{fig:M1500} shows the redshift and absolute magnitude distributions of the selected sample. The median redshift is $z\sim1.9$, and the sample reaches a faint-end magnitude limit of $M_{\mathrm{UV}} = -15.5$. 

\section{Completeness analysis}
\label{sec:comp}

To derive an accurate UV LF, it is essential to estimate the completeness of the galaxy sample and apply appropriate corrections, particularly at the faint end where detection efficiency declines. Following the methodology of \cite{Sun.2026}, we perform completeness simulations by injecting artificial galaxies directly into the real images. For these simulations, we employ a modified version of the GLACIAR2 pipeline \citep{Leethochawalit.2022n}, analogous to the GLACIAR2+ approach developed for lensing fields by \cite{Sun.2026} but adapted here for blank-field observations.

The mock galaxies are generated using the JADES Extragalactic Ultra-deep Artificial Realization (JAGUAR) spectral templates \citep{Williams.2018apjs}, which assume a flat $\Lambda$CDM cosmology. To maintain internal consistency between the completeness simulations and the LF analysis, we use the same cosmological parameters throughout this paper, as already stated in Section \ref{sec:intro}. While we adopt the JAGUAR templates for our completeness simulations, we note that alternative mock catalogs (e.g., the Santa Cruz SAM from \citealt{Yung.2023}) are available for assessing potential systematic uncertainties.

We update the intrinsic size distribution of mock galaxies following the size-luminosity relation from \cite{Shibuya.2015}, extrapolated to luminosities fainter than $M_{\mathrm{UV}} = -16$ when necessary, as in \cite{Sun.2026}. Each artificial galaxy is assigned a S\'ersic profile with index $n = 1.5$, consistent with the findings for star-forming galaxies by \cite{Shibuya.2015}, and is then inserted into the F606W and F435W science images.

\begin{figure}
\centering
\includegraphics[width=0.45\textwidth]{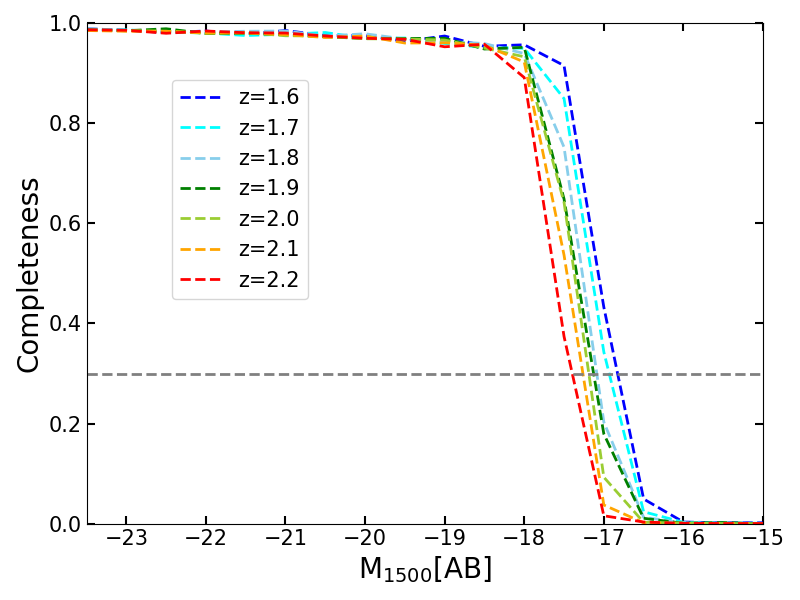}
\caption{Completeness as a function of absolute magnitude $M_{1500}$, with color-coded curves representing different redshifts. The horizontal gray dashed line indicates our $30\%$ completeness threshold.}
\label{fig:C_M}
\end{figure}

The images are then processed through the same SExtractor-based source detection and photometry pipeline as the real data. The completeness function $C(M, z)$ is derived as the fraction of recovered mock galaxies as a function of absolute magnitude and redshift. Figure~\ref{fig:C_M} presents the resulting completeness as a function of $M_{1500}$ for different redshifts. At the bright end ($M_{1500} \lesssim -18$), the completeness approaches $\sim 100\%$, indicating robust recovery of luminous sources. The completeness declines toward fainter magnitudes as the signal-to-noise ratio decreases. Following Paper I for the UV LF of UVCANDELS F275W to avoid large completeness corrections, we apply a $30\%$ completeness threshold (marked by the horizontal gray dashed line in Figure~\ref{fig:C_M}) to define our final sample.

To quantify the systematic uncertainty introduced by the assumed S\'ersic profile, we repeated the completeness simulations with $n=1$ and $n=2.5$, bracketing the typical range for UV-selected star-forming galaxies. The resulting completeness differences relative to the fiducial $n=1.5$ model are largest near the $30\%$ completeness limit at the faint end of our fitting range, where the median differences across redshift are $\Delta C \lesssim 0.04$ for $n=1$ and $\Delta C \lesssim 0.11$ for $n=2.5$, corresponding to relative differences of $\sim7\%$ and $\sim21\%$, respectively.

\section{Luminosity Function}
\label{sec:lf}

To model the galaxy UV LF, we adopt the standard Schechter function parameterization \citep{Schechter76}. In terms of absolute magnitude, the Schechter function is given by
\begin{equation}
\phi(M) = 0.4\ln(10)\phi^{*}10^{-0.4(M - M^{*})(1 + \alpha)}e^{-10^{-0.4(M - M^{*})}},
\label{eq:schechter}
\end{equation}
\noindent where $\alpha$ is the power-law slope at the faint end, $M^{*}$ denotes the characteristic magnitude at which the LF exhibits an exponential cutoff, and $\phi^{*}$ is the normalization factor.

Following the methodology of Paper I, we employ an unbinned maximum likelihood estimate (MLE) to determine the Schechter parameters. This approach maximizes the joint likelihood function constructed as the product of individual detection probabilities, incorporating the effective volume $V_{\mathrm{eff}}(M)$ to correct for survey completeness and a Gaussian distribution to account for magnitude uncertainties. The absolute magnitude uncertainty $\sigma_i$ for each galaxy is the quadrature sum of the photometric error $\sigma_m$ and the redshift estimation error $\sigma_z$, consistent with Paper I. We quantify the uncertainties in our best-fit parameters using Markov Chain Monte Carlo (MCMC) analysis, implemented through the Python package \texttt{emcee} \citep{Foreman13}. We perform the MCMC fitting with 32 chains, each run for 3000 iterations. The first 500 iterations are discarded as burn-in, and the remaining samples are thinned by a factor of 15. The normalization parameter $\phi^{*}$, which cancels out in the MLE fit, is estimated separately from the number counts.

\begin{figure*}
\centering
\includegraphics[width=0.7\textwidth]{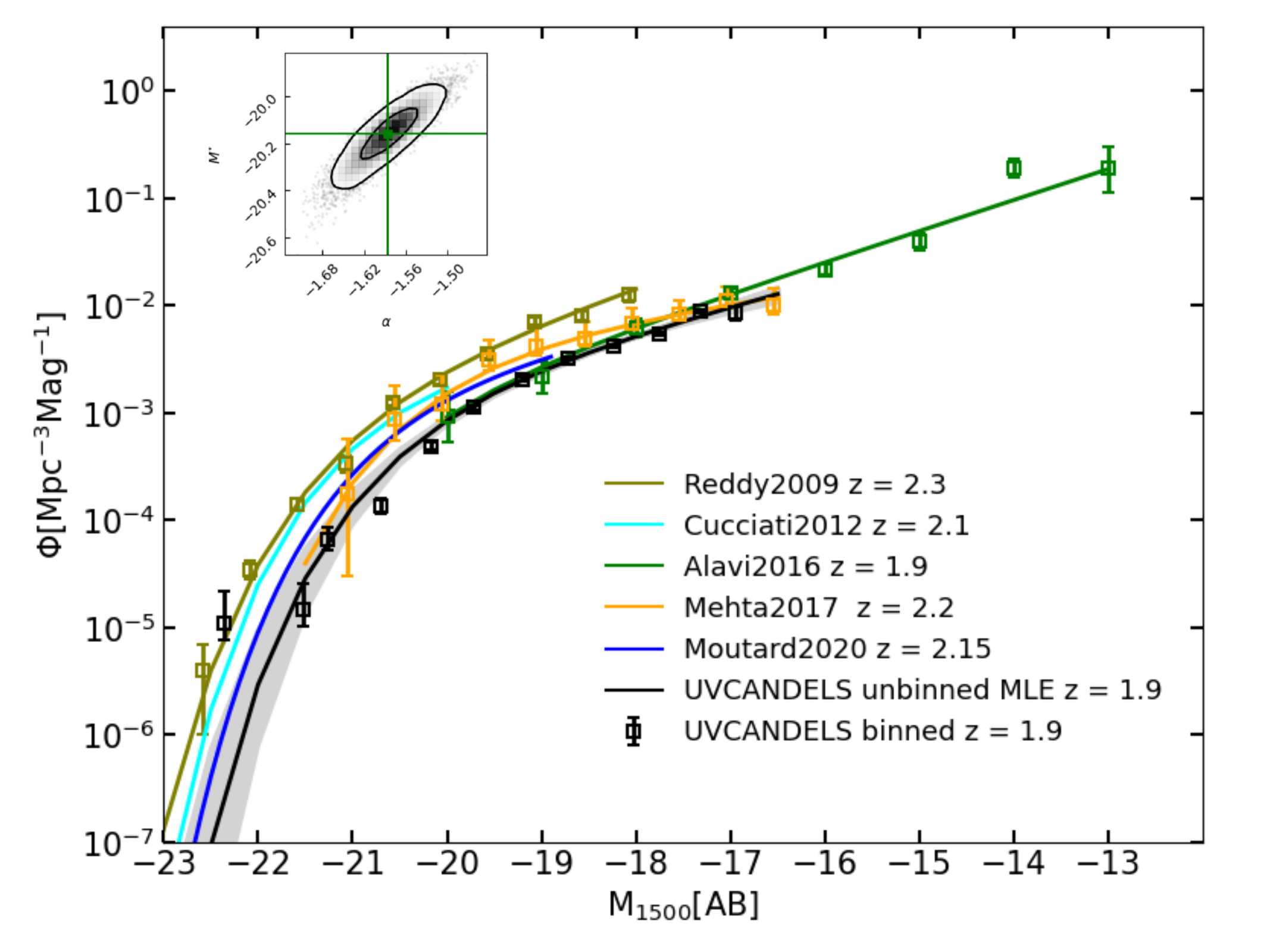}
\caption{Rest-frame UV LF of UVCANDELS galaxies at $1.6 < z < 2.2$ (median $z\sim1.9$). The black solid line shows the best-fit Schechter function derived from the unbinned MLE, where each galaxy contributes individually to the joint likelihood. The gray shaded region represents the $3\sigma$ uncertainties from the MLE. The black squares represent the binned LF for visual inspection only. The colored lines indicate results from the literature at similar redshifts, as summarized in the legend. The inset shows the MCMC posterior distributions for the Schechter LF parameters $\alpha$ and $M^{*}$. The green cross marks the best-fit (median) values, reported in Table~\ref{tab:pars}, and the black contours indicate the $1\sigma$ ($0.393$) and $2\sigma$ ($0.865$) confidence levels.}
\label{fig:lf}
\end{figure*}

Figure~\ref{fig:lf} presents the rest-frame UV LF for our galaxy sample at a median redshift of $z \sim 1.9$, derived using the MLE technique described above. After applying the $30\%$ completeness threshold, our final sample consists of 5,212 galaxies. The best-fit Schechter function is shown as the black solid curve, with the surrounding gray shaded band representing the $3\sigma$ confidence region evaluated from the MCMC chains. The inset panel presents the joint constraints on the Schechter parameters, showing the $1\sigma$ and $2\sigma$ confidence contours for $\alpha$ and $M^{*}$, with the green lines marking the best-fit (median) values. 

We complement the unbinned maximum likelihood fit with a binned representation of the LF (black squares), which includes Poissonian uncertainties and serves only for visual comparison. The binned LF is computed following Paper I, using the $1/V_{\mathrm{eff}}$ estimator. The error bars for the binned data are derived from Poisson statistics, with appropriate corrections applied to bins containing fewer than 50 galaxies following the formulation of \cite{Gehrels86}. All bins have a fixed width of $\Delta M_{\mathrm{UV}} = 0.5$ mag. 

\begin{deluxetable*}{lllllll}
\tabletypesize{\footnotesize}
\tablecaption{Best-fit Schechter Parameters for UV LFs and the UV Luminosity Density}  
\tablehead{ 
\colhead{Redshift} & \colhead{$M_{lim,\mathrm{UV}}(\mathrm{Mag})$} & \colhead{$N$\tablenotemark{a}} & \colhead{$\alpha$} & \colhead{$M^{*}(\mathrm{Mag})$} & \colhead{$\phi^{*} (10^{-3}\mathrm{Mpc}^{-3})$} & \colhead{$\rho_{\mathrm{UV}}$\tablenotemark{b}}
}
\startdata
\multicolumn{7}{c}{Full Sample}\\
\hline
$1.6<z<2.2$ & -16.83 & 5212 & $-1.586^{+0.040}_{-0.042}$ & $-20.160^{+0.101}_{-0.117}$ & $2.01^{+0.30}_{-0.30}$  & $1.821^{+0.032}_{-0.037}$ \\
\hline
\multicolumn{7}{c}{Sub-samples}\\
\hline
$1.6<z<1.9$ & -16.83 & 2827 & $-1.488^{+0.055}_{-0.054}$ & $-19.875^{+0.123}_{-0.131}$ & $2.89^{+0.50}_{-0.47}$ & $1.717^{+0.036}_{0.039}$ \\
$1.9<z<2.2$ & -17.14 & 2482 & $-1.604^{+0.066}_{-0.062}$ & $-20.420^{+0.196}_{-0.204}$ & $1.55^{+0.46}_{-0.37}$  & $1.877^{+0.066}_{-0.063}$
\enddata
\label{tab:pars}
\tablenotetext{a}{Sample size after removing sources with completeness $<30\%$.}
\tablenotetext{b}{in units of $\times 10^{26}$ ergs/s/Hz/Mpc$^3$}
\end{deluxetable*}

Figure~\ref{fig:lf} compares our results with published rest-frame UV LFs at $\sim1500~\AA$ and similar redshifts, including data from VVDS \citep{Cucciati12}, HFF lensing fields \citep{Alavi16}, UVUDF \citep{Mehta.2017}, and CLAUDS/HSC-SSP \citep{Moutard20}. Our maximum likelihood analysis yields a faint-end slope of $\alpha = -1.586 \pm 0.042$, a characteristic magnitude of $M^{*} = -20.160 \pm 0.117$, and a normalization of $\phi^{*} = (2.01 \pm 0.30) \times 10^{-3}$ Mpc$^{-3}$ at $z \sim 1.9$. These best-fit parameters are also summarized in Table~\ref{tab:pars}. Our measured faint-end slope is consistent with that of \cite{Alavi16} ($\alpha = -1.72 \pm 0.04$ at $z \sim 1.9$) at the $2.3\sigma$ level, while being steeper than the measurement of \cite{Mehta.2017} ($\alpha = -1.32 \pm 0.14$ at $z \sim 2.2$). 

We note that cosmic variance contributes an additional uncertainty to our measurements, but it is subdominant compared to the Poisson errors at the magnitude limits probed by our sample. To quantify this contribution, we employ the \texttt{Cosmic Variance Calculator v1.03} \citep{Trenti08}, implementing the halo bias model of \cite{Sheth99} with parameter choices $\sigma_8 = 0.8$ and an average halo occupation fraction of $0.5$. For galaxies in our brightest magnitude bins, the fractional error introduced by cosmic variance is estimated to be approximately $0.34$, which is approximately one-third of the Poisson uncertainty.

\section{Discussion}
\label{sec:discuss}

\subsection{Redshift Evolution of the UV LF}
\label{sec:evol}

\begin{figure*}
\centering

\includegraphics[width=\columnwidth]{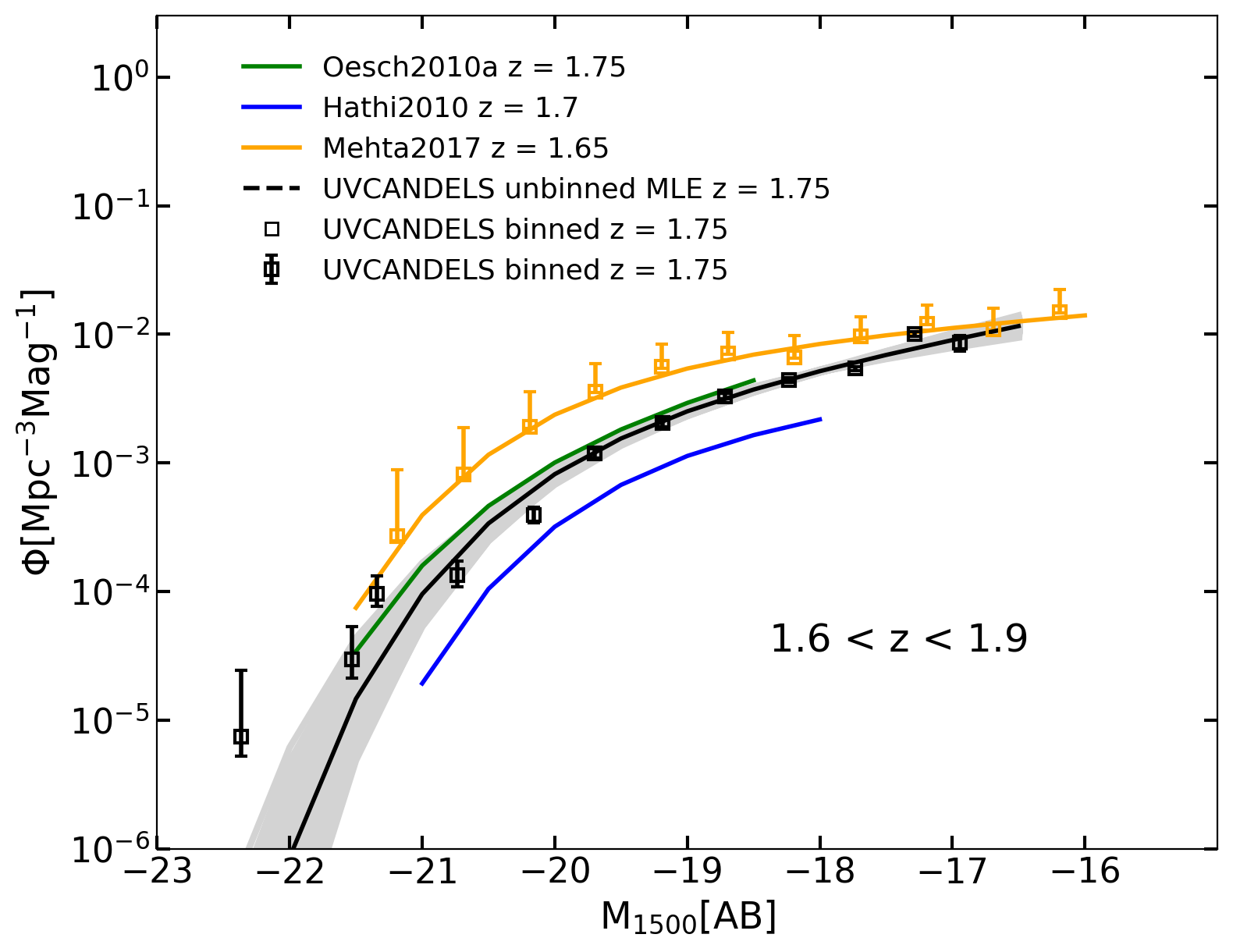}
\includegraphics[width=\columnwidth]{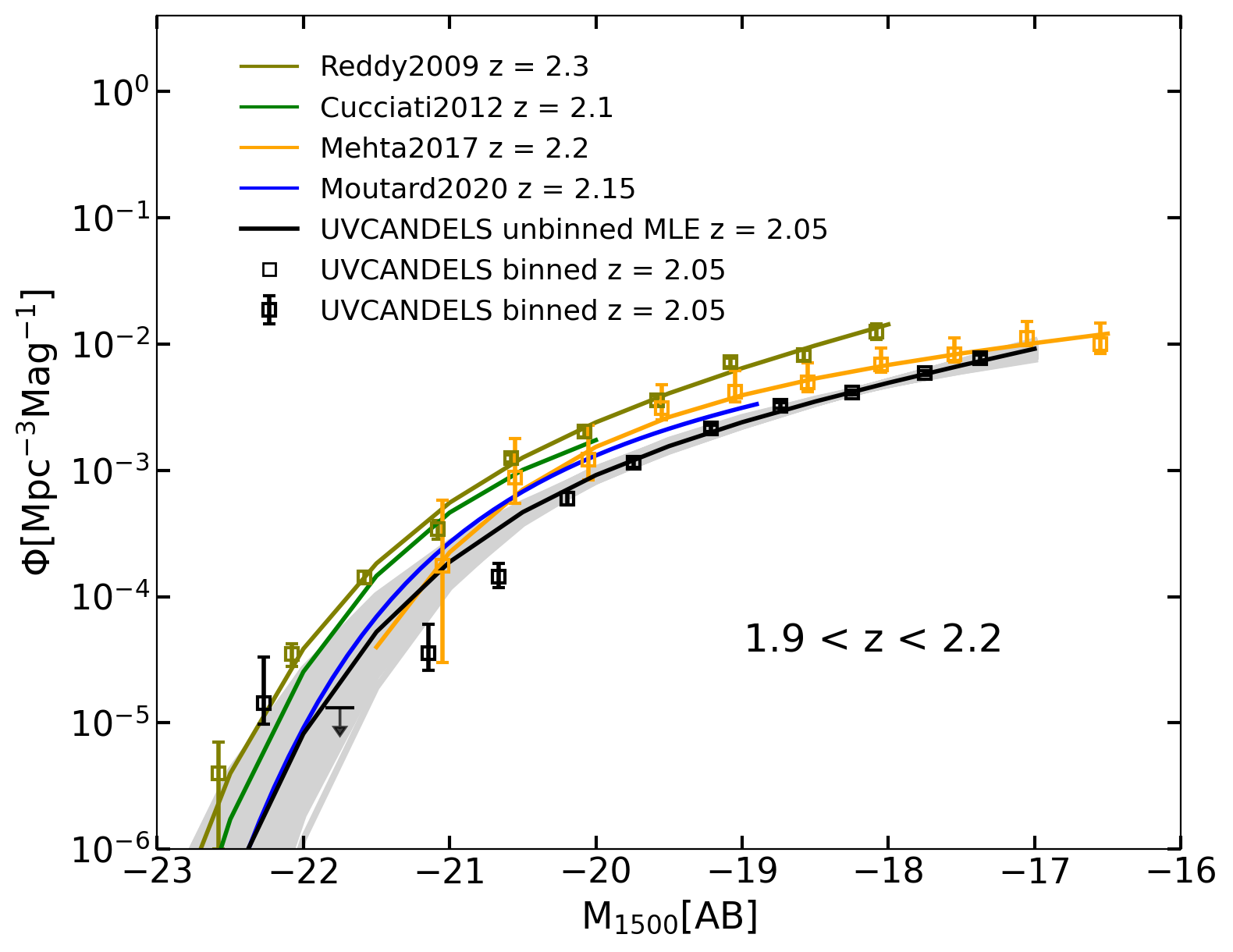}
\caption{Rest-frame UV LFs of the split sub-samples at $1.6<z<1.9$ (left panel) and $1.9<z<2.2$ (right panel). Symbols and curves follow the same conventions as in Figure \ref{fig:lf}. Note that in the left panel, the three brightest bins all have non-zero counts; the apparent gap between the brightest bin and the next two arises because we use the median magnitude to represent each bin. In the right panel, the second brightest bin has zero counts and is shown as an upper limit.}
\label{fig:lf-spliting}
\end{figure*}

To investigate the redshift evolution of the UV LF, we divide our full sample into two redshift bins: a lower-redshift bin ($1.6 < z < 1.9$) and a higher-redshift bin ($1.9 < z < 2.2$). The best-fit Schechter parameters for each subsample are listed in Table~\ref{tab:pars}. Figure~\ref{fig:lf-spliting} presents the rest-frame UV LFs for the two redshift bins, respectively. Comparing the two redshift bins, we observe a mild steepening in the faint-end slope $\alpha$ from $-1.488\pm0.055$ at $z\sim1.75$ to $-1.604\pm0.066$ at $z\sim2.05$. Considering this is only a $\sim1.4\sigma$ difference, the evolution is statistically insignificant. Meanwhile, the characteristic magnitude $M^{*}$ brightens from $-19.875\pm0.131$ to $-20.420\pm0.204$ ($\sim2.3\sigma$ difference). At the bright end of these UV LFs, the small number of galaxies per bin (typically 0–3) leads to large Poisson fluctuations, which naturally explains the presence of empty bins and the apparent jumps between adjacent bins. In the right panel, the two sources in the brightest bin have been rigorously checked against AGN and stellar catalogs, with no significant anomalies found upon visual inspection. The second brightest bin has zero counts and we accordingly plot an upper limit. Note that in the left panel, the apparent gap between the brightest bin and the next two is a visual artifact of using median magnitudes (see also the caption). 

In comparing our results with the literature, we note that most previous studies assume a flat $\Lambda$CDM cosmology with $\Omega_m = 0.3$, $\Omega_\Lambda = 0.7$, and $H_0 = 70$ km s$^{-1}$ Mpc$^{-1}$, identical to the cosmology adopted in this work. While a small number of studies adopted different cosmologies, Paper I, for instance, used the Planck 2020 cosmology ($\Omega_m = 0.315$, $\Omega_\Lambda = 0.685$, $H_0 = 67.4$) at $z \sim 0.8$ but found that the resulting differences in the best-fit Schechter parameters were comparable to their $1\sigma$ statistical uncertainties. We therefore ignore these minor cosmological differences and compile literature results together to discuss the redshift evolution of the Schechter parameters. Regarding the rest-frame wavelength, \citealt{Bouwens.2022c} measured the UV LF at 1600~\AA, while most other works (including ours) adopt 1500~\AA. However, the resulting systematic offset is smaller than our statistical uncertainties.

\begin{figure*}
\centering
\includegraphics[width=0.7\textwidth]{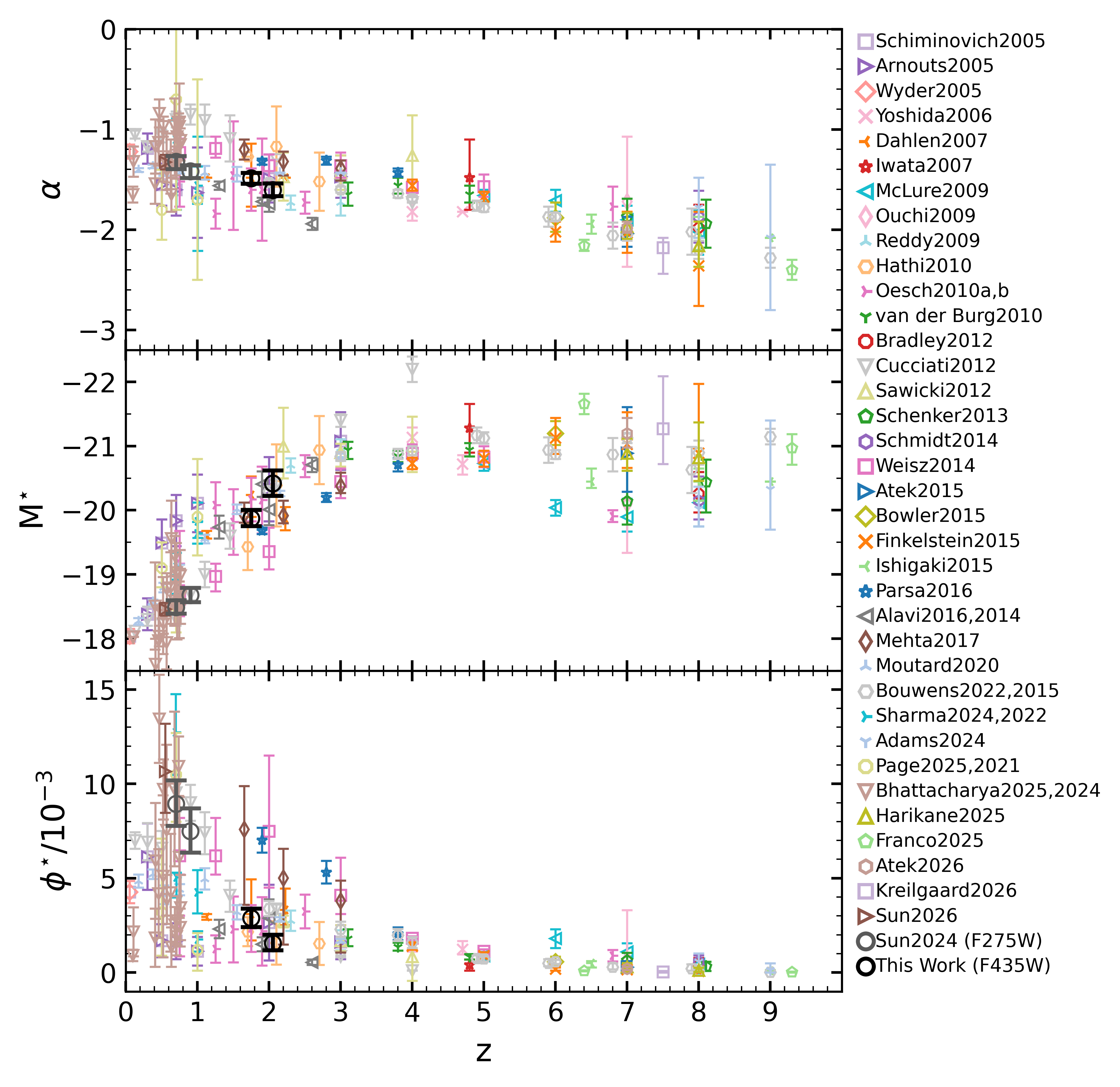}
\caption{Redshift evolution of the Schechter parameters of the UV LF. The black circles show the results from the UVCANDELS F435W sample at $z\sim1.75$ and $2.05$. The gray circle highlights our previous results from UVCANDELS F275W. Symbols representing previous determinations from the literature are summarized in the legend. The error bars indicate $1\sigma$ uncertainties.}
\label{fig:par_evol}
\end{figure*}

Figure~\ref{fig:par_evol} compiles measurements of the Schechter parameters ($\alpha$, $M^{*}$, and $\phi^{*}$) from UVCANDELS (this work and paper I) and from the literature over a broad redshift range of $0 < z < 10$ \citep{Schiminovich.2005, Arnouts05, Wyder.2005, Yoshida.2006, Dahlen.2007, Iwata.2007, Mclure.2009, Ouchi.2009, Reddy.2009, Hathi.2010, Oesch10, Oesch.2010L21, Vanderburg.2010, Bradley.2012, Cucciati12, Sawicki.2012, Schenker.2013, Schmidt.2014, Weisz.2014, Atek.2015,Bowler.2015, Finkelstein.2015LF, Ishigaki.2015, Parsa.2016, Alavi16, Alavi14, Mehta.2017, Moutard20, Bouwens.2022c, Bouwens.2015, Sharma.2024, Sharma.2022, Adams.2024, Page.2025, Page.2021, Bhattacharya.2025, Bhattacharya.2024, Harikane.2025, Franco.2025, Atek.2026, Kreilgaard.2026, Sun.2024, Sun.2026}. 
To quantify the redshift evolution of the faint-end slope $\alpha$ across the full redshift range probed by UVCANDELS, we perform a linear fit to the four measurements from both F275W and F435W samples. The fit yields $d\alpha/dz = -0.164 \pm 0.041$, with a reduced chi-square of $\chi^2_\nu =0.586$. This corresponds to a $\sim4.0\sigma$ deviation from a constant $\alpha$, suggesting a significant steepening of the faint-end slope with increasing redshift, in good agreement with the literature compilation. Our observed steepening of $\alpha$ at higher redshifts is also qualitatively consistent with the evolutionary trends found in the stellar mass function \citep[e.g.,][]{Conselice.2016}. This steepening indicates that fainter galaxies become increasingly dominant relative to brighter ones at higher redshifts, reflecting that low-mass star-forming galaxies contribute more to the total UV luminosity density at early epochs. Specifically, our $\alpha$ values at $z\sim1.75$ and $z\sim2.05$ are consistent with those from \cite{Reddy.2009}, \cite{Oesch10}, and \cite{Alavi16} at similar redshifts, all falling within $3\sigma$ uncertainties. The evolution of $M^{*}$ shows a steady brightening from $z \sim 0$ to $z \sim 2.5$, indicating that typical star-forming galaxies were more luminous at earlier cosmic times, likely due to more intense star formation and lower dust attenuation. Our $M^{*}$ values at both redshift bins fall neatly along this trend, in good agreement with \cite{Hathi.2010}, \cite{Oesch10} and \cite{Alavi16} at comparable epochs. Overall, our results reinforce the picture that galaxies at higher redshifts have steeper faint-end slopes, brighter characteristic magnitudes, and lower overall number densities across cosmic history up to $z \sim 10$.

\subsection{UV Luminosity Density}
%\label{sec:uvld}

The total unobscured (i.e., no dust correction) UV light emitted by galaxies, quantified by the cosmic UV luminosity density $\rho_\text{UV}$, is derived from our best-fit Schechter function (Section~\ref{sec:lf}). This quantity is calculated by integrating the product of luminosity and the LF over the full range of galaxy brightness:

\begin{equation}
\label{eq:uvld}
\rho_\text{UV} = \int_{L_\text{lim}}^{\infty} L\ \phi(L)\ dL= \int_{-\infty}^{M_\text{lim}} L(M)\ \phi(M)\ dM
\end{equation}

In Table~\ref{tab:pars}, we present the cumulative UV luminosity density computed by integrating down to $M_\text{UV}=-15$. This limit is motivated by previous studies (e.g., \cite{Bouwens.2022c, Sun.2026} and references therein), which rule out a turnover in the UV LF brighter than $M_{\mathrm{UV}}\sim -15$ (at the $\sim$2--3$\sigma$ level) over most of the redshift range. The uncertainty in $\rho_\mathrm{UV}$ is propagated from the fitted Schechter parameters ($\alpha$, $M^{*}$). We achieve this by first drawing a large sample of $(\alpha, M^{*})$ pairs from their two-dimensional posterior distribution via MCMC. For each sampled pair, we evaluate Equation~(\ref{eq:uvld}) to compute a corresponding $\rho_\mathrm{UV}$ value. The ensemble of these computed values forms a probability distribution for $\rho_\mathrm{UV}$, from which we directly infer its confidence interval.
 
\begin{figure*}
\centering
\includegraphics[width=0.7\textwidth]{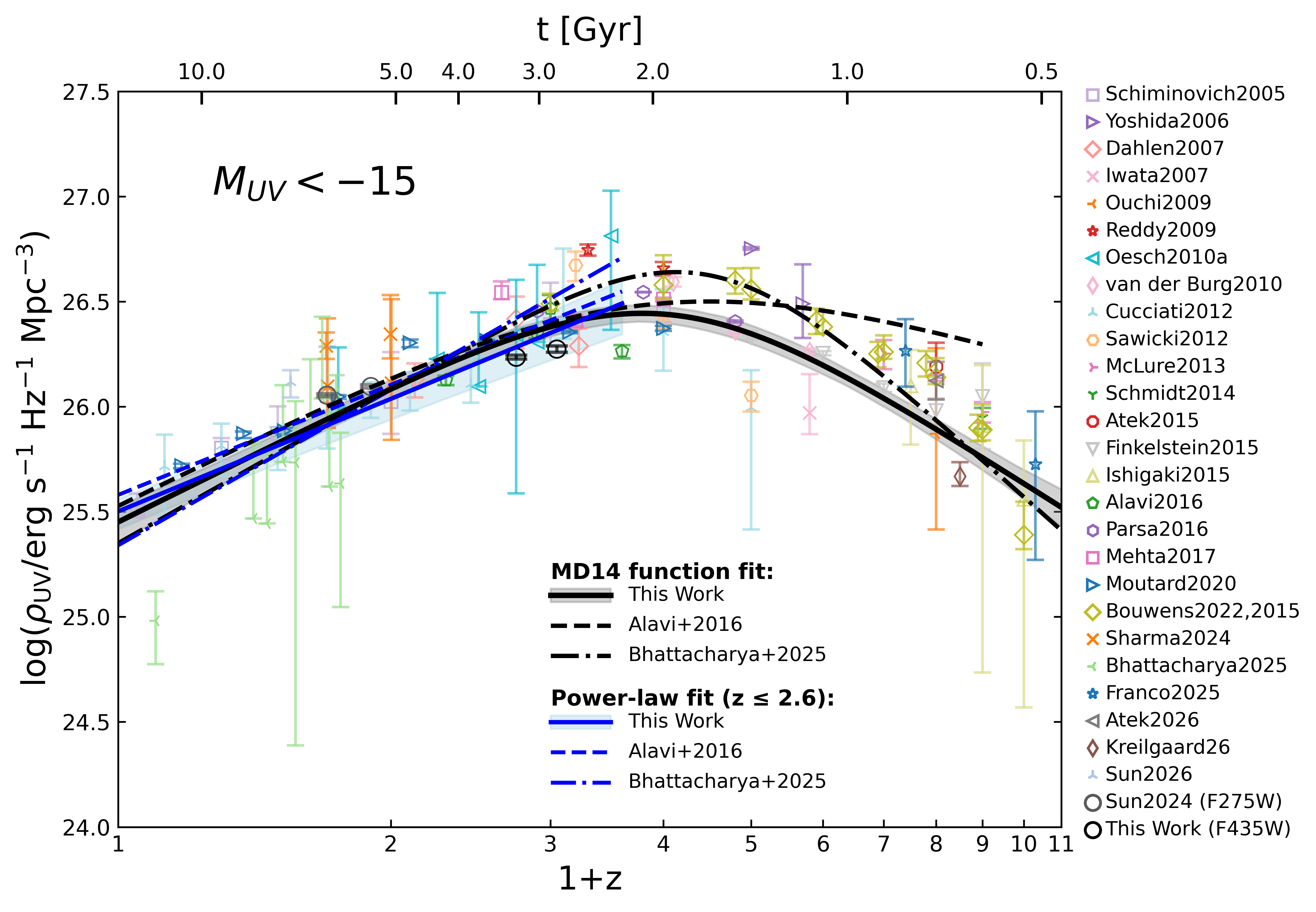}
\caption{Redshift evolution of the unobscured (no dust correction) UV luminosity density. The black open circles show the results derived from the best-fit Schechter parameters of our sample at $z\sim1.75$ and $z\sim2.05$. The gray square highlights our previous Paper I results from UVCANDELS F275W. Literature results are summarized in the legend. All data points are computed by integrating the rest-frame UV LFs down to $M_{\mathrm{UV}} = -15$. Our best-fit MD14 function and the power-law fit for $z \leq 2.6$ are shown as the black and blue solid lines, respectively, with shaded regions representing the $3\sigma$ uncertainties. or comparison, we also plot the results from \citet[][dashed lines]{Alavi16} and \citet[][dash-dotted lines]{Bhattacharya.2025}.}
\label{fig:rho_evol}
\end{figure*}

Figure~\ref{fig:rho_evol} presents the evolution of the UV luminosity density over the redshift range $0<z<10$. To ensure a uniform comparison, we have recomputed every literature $\rho_\text{UV}$ value by applying Equation~(\ref{eq:uvld}) with the Schechter parameters reported in the original studies, integrating consistently down to our adopted limit of $M_\text{UV} = -15$ \citep{Schiminovich.2005, Yoshida.2006, Dahlen.2007, Iwata.2007, Ouchi.2009, Reddy.2009, Oesch10, Vanderburg.2010, Cucciati12, Sawicki.2012, McLure.2013, Schmidt.2014, Atek.2015, Finkelstein.2015LF, Ishigaki.2015, Alavi16, Parsa.2016, Mehta.2017, Moutard20, Bouwens.2022c, Sharma.2024, Bhattacharya.2025, Franco.2025, Atek.2026, Kreilgaard.2026, Sun.2024, Sun.2026}. Note that uncertainty estimation of $\rho_\mathrm{UV}$ cannot be done straightforwardly, since covariance between Schechter parameters are usually not provided in the literature. Following \cite{Madau.2014}, we consequently employ fixed fractional uncertainties ($\Delta\rho_\mathrm{UV}/\rho_\mathrm{UV}$) reported in the original studies to derive consistent error estimates for $\rho_\mathrm{UV}$ values integrated to $M_\text{UV}=-15$. 

\cite{Alavi16} and \cite{Bhattacharya.2025} performed power-law fits to the $\rho_{\mathrm{UV}}$ evolution over $z<2.6$, i.e., around and after the peak of cosmic star formation (cosmic noon). Since our UVCANDELS measurements also primarily fall within this redshift range ($z \lesssim 2$), we carry out the same power-law fit for consistency and direct comparison. Using the same data selection, we obtain $\log_{10}(N) = 25.50 \pm 0.027$ and $\beta = 1.77 \pm 0.07$ for the power-law function $\rho_{\mathrm{UV}} = N(1+z)^{\beta}$, where $N$ is in units of $\mathrm{erg~s^{-1}~Hz^{-1}~Mpc^{-3}}$. For comparison, \cite{Alavi16} reported $\log_{10}(N) = 25.58$ and $\beta = 1.74$, while \cite{Bhattacharya.2025} obtained $\log_{10}(N) = 25.34 \pm 0.06$ and $\beta = 2.46 \pm 0.2$. Our power-law parameters are broadly consistent with those of \cite{Alavi16}, but are significantly shallower than the slope reported by \cite{Bhattacharya.2025} ($\sim3.3\sigma$).

To model the redshift evolution of $\rho_{\mathrm{UV}}$ over a broader baseline up to $z \sim 10$, we adopt the modified double power-law function originally introduced by \cite{Madau.2014} for the cosmic star formation rate density and later adapted by \cite{Alavi16} for UV luminosity density (hereafter MD14):

\begin{equation}
\rho_{\mathrm{UV}} = a \times \frac{(1+z)^b}{1 + [(1+z) / d]^c} \ 10^{26}\ \mathrm{erg\ s^{-1}\ Hz^{-1}\ Mpc^{-3}},
\label{eq:md14}
\end{equation}

\noindent where $a$ is the normalization, $b$ and $c$ characterize the growth and decline slopes, respectively, and $d$ controls the peak location. We combine our UVCANDELS F435W ($z \sim 1.6-2.2$) and F275W ($z \sim 0.6-1$) measurements with literature data at lower and higher redshifts to constrain the global evolution of $\rho_{\mathrm{UV}}$. Following \cite{Alavi16}, we add a $12\%$ systematic uncertainty in quadrature to our $\rho_{\mathrm{UV}}$ measurements when performing the power-law and MD14 fits, to account for potential systematics at faint luminosities.

The best-fit MD14 function is shown in Figure~\ref{fig:rho_evol} as the black solid curve, with parameters $a = 0.28 \pm 0.02$, $b = 2.17 \pm 0.13$, $c = 4.82 \pm 0.12$, and $d = 3.95 \pm 0.19$ (the power-law fit for $z \leq 2.6$ is shown as the blue solid line). As illustrated in the figure, the MD14 functions from \cite{Alavi16}, \cite{Bhattacharya.2025}, and this work are broadly consistent at $z \lesssim 2$. At higher redshifts, however, noticeable differences appear. In particular, the \cite{Alavi16} fit deviates more significantly, likely due to the limited availability of high-redshift data in their fitting sample. Our best-fit MD14 parameters imply that the unobscured (no dust correction) UV luminosity density peaks at $z = 2.79 \pm 0.21$, which is earlier than the peak of the total (dust-corrected) star formation rate density at $z \sim 1.86$ found by \cite{Madau.2014}. This offset suggests that dust obscuration becomes increasingly important from higher redshifts toward the peak epoch of cosmic star formation at $z\sim2$, where a significant fraction of star formation is hidden from UV view.

In the coming years, facilities such as Roman and Euclid will further advance the field, while JWST continues to push the high-redshift frontier. Among these, the Chinese Space-station Survey Telescope (CSST) is particularly well-suited to improve UV LF and $\rho_{\rm UV}$ measurements at $z \lesssim 1$ up to $z\sim6$. Building upon the UV-optimized aperture photometry method developed in Paper I, CSST deep fields ($\sim400$ deg$^2$) will achieve significantly greater depth in the NUV (reaching $\gtrsim27$ AB) and blue optical bands, precisely probing the UV LF faint end at $z\lesssim2$. Taking advantage of lensing clusters within the same deep fields, CSST will reach $M_{\rm UV}\sim-12$, directly constraining the faint end into the regime of potential turnover. At $z\sim3-6$, CSST will extend $\rho_{\rm UV}$ measurements into the reionization epoch, helping to determine whether faint star-forming galaxies dominate cosmic reionization.

\section{Summary}
\label{sec:sum}

In this paper, we presented an analysis of the rest-frame UV LF at $1.6<z<2.2$ using the UVCANDELS F435W imaging data. Taking advantage of the UV-optimized aperture photometry method, well-tested photometric redshifts, and a careful completeness correction, we derived the Schechter parameters of the UV LF via maximum likelihood estimation.

We constructed a catalog of 5,212 galaxies at $1.6<z<2.2$ down to $M_{\mathrm{UV}} = -16.8$ after applying a $30\%$ completeness threshold, with a median redshift of $z\sim1.9$. The rest-frame UV LF at $z\sim1.9$ is well described by a Schechter function with a faint-end slope $\alpha = -1.586^{+0.040}_{-0.042}$, a characteristic magnitude $M^{*} = -20.160^{+0.101}_{-0.117}$, and a normalization $\phi^{*} = 2.01^{+0.30}_{-0.30}\times10^{-3}$ Mpc$^{-3}$. Our measured $\alpha$ is consistent with the value from \cite{Alavi16} at $z\sim1.9$ at the $2.3\sigma$ level, and is steeper than the measurement reported by \cite{Mehta.2017} at $z\sim2.2$.

By splitting the full sample into two redshift bins ($1.6<z<1.9$ and $1.9<z<2.2$), we investigated the redshift evolution of the UV LF parameters. The faint-end slope $\alpha$ exhibits a mild steepening from $-1.488\pm0.055$ at $z\sim1.75$ to $-1.604\pm0.066$ at $z\sim2.05$. Given the $\sim1.4\sigma$ difference, this variation is not statistically significant. The characteristic magnitude $M^{*}$, meanwhile, brightens from $-19.875\pm0.131$ to $-20.420\pm0.204$ ($\sim2.3\sigma$ difference). Combining the F275W measurements from Paper I with the F435W results from this work, we examine the redshift evolution of the faint-end slope across $0.6<z<2.2$. A weighted linear fit to the four $\alpha$ values yields $d\alpha/dz = -0.164 \pm 0.041$ ($\chi^2_\nu = 0.586$), corresponding to a $\sim4.0\sigma$ deviation from no evolution. These trends are broadly consistent with literature results.

Integrating our best-fit Schechter function down to $M_{\mathrm{UV}} = -15$, we derived the unobscured UV luminosity density $\rho_{\mathrm{UV}} = 1.821^{+0.032}_{-0.037}\ (\times10^{26}\ \mathrm{erg\ s^{-1}\ Hz^{-1}\ Mpc^{-3}})$ at $z\sim1.9$. Combined with our previous F275W result at $z\sim0.8$ in Paper I and literature measurements, we modeled the redshift evolution of $\rho_{\mathrm{UV}}$ using a power-law fit at $z\leq2.6$ and the MD14 function over $0<z<10$. The power-law fit yields $\log_{10}(N) = 25.50 \pm 0.027$ and $\beta = 1.77 \pm 0.07$, broadly consistent with those of \cite{Alavi16} but differing from the steeper slope reported by \cite{Bhattacharya.2025}. The MD14 fit gives parameters $a = 0.28 \pm 0.02$, $b = 2.17 \pm 0.13$, $c = 4.82 \pm 0.12$, and $d = 3.95 \pm 0.19$. Our MD14 function is broadly consistent with those from \cite{Alavi16} and \cite{Bhattacharya.2025} at $z\lesssim2$, but deviates at higher redshifts, likely due to the limited high-redshift data especially in the \cite{Alavi16} sample.

Overall, combined with our previous UVCANDELS F275W results \citep{Sun.2024}, the wide-field deep UV and blue optical imaging from UVCANDELS provide robust constraints on the faint-end slope of the UV LF from $z\sim0.5$ up to cosmic noon. The steepening of $\alpha$ with redshift and the rising trend of $\rho_{\mathrm{UV}}$ toward $z\sim2$ highlight the important role of low-luminosity galaxies in the global star formation budget during the peak epoch of cosmic star formation. The UVCANDELS F435W photometric catalogs presented in this work have been made publicly available on MAST\citep{UVCANDELS_MAST}.

%\begin{acknowledgments}
\section*{ }
\noindent We thank the anonymous referee for constructive comments that improved the quality of this paper.
We thank Yuxuan Pang and Shengzhe Wang for helpful discussion. 
This work is supported by Beijing Natural Science Foundation (QY26525), the National Natural Science Foundation of China (grant 12373009), the China Manned Space Program with grant no. CMS-CSST-2025-A06, the National Key R\&D Program of China No.2025YFF0510603, the CAS Project for Young Scientists in Basic Research Grant No. YSBR-062, and the Fundamental Research Funds for the Central Universities. 
XW acknowledges the support by the Xiaomi Young Talents Program, and the work carried out, in part, at the Swinburne University of Technology, sponsored by the ACAMAR visiting fellowship. 
RAW acknowledges support from NASA JWST Interdisciplinary Scientist grants NAG5-12460, NNX14AN10G and 80NSSC18K0200 from GSFC.
This work is based on observations with the NASA/ESA Hubble Space Telescope obtained at the Space Telescope Science Institute, which is operated by the Association of Universities for Research in Astronomy, Incorporated.
%\end{acknowledgments}

%\bibliographystyle{aasjournal}
%\bibliography{xinlib,reflei,newref}

\end{document}